\documentclass[a4paper,fleqn]{cas-dc}

\usepackage{amsmath,amsfonts}
\usepackage{algorithmic}
\usepackage{algorithm}
\usepackage{array}
\usepackage{textcomp}
\usepackage{stfloats}
\usepackage{url}
\usepackage{verbatim}
\usepackage{graphicx}
\usepackage{cite}
\usepackage{pifont}
\newcommand{\cmark}{\ding{51}}%
\newcommand{\xmark}{\ding{55}}

\usepackage{hhline}
\usepackage{adjustbox}
\usepackage{float}
\usepackage{booktabs}
\usepackage{multirow}
\usepackage{tabu}
\usepackage{lscape}
\usepackage[normalem]{ulem}

\usepackage{arydshln}

\usepackage{lettrine}
\usepackage{verbatim}

\usepackage{url}
\usepackage{lscape}
\usepackage{comment, soul}
\usepackage{bm}
\usepackage{subcaption}
\usepackage{diagbox}
\usepackage{slashbox}
\usepackage{makecell}
\usepackage{xspace}
\usepackage{acronym}

\newcommand{\etal}{\emph{et al.}\xspace}

\newcommand{\ie}{\emph{i.e.}, }

\usepackage{caption}
\usepackage[edges]{forest}
\usepackage{tikz}
\usetikzlibrary{shadows.blur}
\usepackage{cite}
\usepackage{paralist}
\usepackage{amsmath,amssymb,amsfonts}
\usepackage{algorithmic}
\usepackage{graphicx}
\usepackage{longtable}
\usepackage{threeparttable}
\usepackage{textcomp}
\usepackage{xcolor}

\def\BibTeX{{\rm B\kern-.05em{\sc i\kern-.025em b}\kern-.08em
    T\kern-.1667em\lower.7ex\hbox{E}\kern-.125emX}}

\usepackage{siunitx}

\usepackage[numbers]{natbib}
\usepackage{epsfig}
\usepackage[inline]{enumitem}
\usepackage{caption}
\usepackage{stackengine}
\usepackage{subcaption}
\usepackage{balance}
\usepackage{xspace}
\usepackage{amsmath}

\newcolumntype{?}{!{\vrule width 1pt}}
\def\tsc#1{\csdef{#1}{\textsc{\lowercase{#1}}\xspace}}
\tsc{WGM}
\tsc{QE}
\tsc{EP}
\tsc{PMS}
\tsc{BEC}
\tsc{DE}
\usepackage{acronym}
\newacro{has}[HAS]{HTTP adaptive streaming}
\newacro{uav}[UAV]{unmanned aerial vehicle}
\newacro{vr}[VR]{virtual reality}
\newacro{xr}[XR]{extended reality}
\newacro{vista}[ODVista]{omnidirectional video streaming dataset}
\newacro{ra}[RA]{random access}
\newacro{sr}[SR]{super-resolution}
\newacro{hevc}[HEVC]{high-efficiency video coding}
\newacro{odv}[ODV]{omnidirectional video}
\newacro{cc}[CC]{creative commons attribution}
\newacro{erp}[ERP]{equirectangular projection}
\newacro{cmp}[CMP]{map projection}
\newacro{si}[SI]{spatial information}
\newacro{ti}[TI]{temporal information}
\newacro{cf}[CF]{colorfulness}
\newacro{vca}[VCA]{video complexity analyzer}
\newacro{br}[BR]{brightness}
\newacro{cnn}[CNN]{convolution neural network}
\newacro{rnn}[RNN]{recurrent neural network}
\newacro{gan}[GAN]{generative adversarial network}
\newacro{vit}[ViT]{vision transformer}
\newacro{lr}[LR]{low resolution}
\newacro{hr}[HR]{high resolution}
\newacro{ra}[RA]{random access}
\newacro{gpu}[GPU]{graphics processing units}
\newacro{ws-psnr}[WS-PSNR]{weighted spherical peak signal-to-noise ratio}
\newacro{ws-ssim}[WS-SSIM]{weighted spherical structural similarity index measure}
\newacro{ssim}[SSIM]{structural similarity index measure}
\newacro{npu}[NPU]{neural processing unit}
\newacro{gpu}[GPU]{graphics processing unit}
\newacro{ai}[AI]{artificial intelligence}
\newacro{fps}[fps]{frames per second}
\newacro{ml}[ML]{machine learning}
\newacro{vod}[VoD]{video-on-demand}
\newacro{qoe}[QoE]{quality of experience}
\newacro{fsrcnn}[FSRCNN]{fast super-resolution convolutional neural network}
\newacro{vvc}[VVC]{versatile video coding}
\newacro{sisr}[SISR]{single image super-resolution}
\newacro{dnn}[DNN]{deep neural network}
\newacro{srcnn}[SRCNN]{super-resolution convolutional neural network}
\newacro{rstb}[RSTB]{residual swin transformer block}
\newacro{psnr}[PSNR]{peak signal-to-noise ratio}
\newacro{lau-net}[AU-Net]{latitude adaptive upscaling network}
\newacro{sliif}[SLIIF]{spherical latent implicit image function}
\newacro{odi}[ODI]{omnidirectional image}
\newacro{hr}[HR]{high-resolution}
\newacro{lr}[LR]{low-resolution}
\newacro{opdb}[OPDB]{omnidirectional position-aware deformable block}

\newacro{csa}[CSA]{cross-scale attention}
\newacro{wt}[WT]{wavelet hallucination}
\newacro{hat}[HAT]{hybrid channel attention}
\newacro{flop}[FLOP]{floating point operation}
\newacro{bd-br}[BD-BR]{Bjøntegaard delta bitrate}
\newacro{dat}[DAT]{dual aggregation transformer}
\newacro{aim}[AIM]{adaptive interaction module}
\newacro{sgfn}[SGFN]{spatial-gate feed-forward network}
\newacro{vsr}[VSR]{video super resolution}
\newacro{resnet}[ResNet]{residual networks}
\newacro{dcn}[DCN]{deformable convolution}
\newacro{grn}[GRN]{Global Response Normalization}
\newacro{safm}[SAFM]{spatially-adaptive feature modulation}
\newacro{ffc}[FFC]{fast Fourier convolution}
\newacro{fft}[FFT]{fast Fourier transform}
\newacro{cspsr}[CSPSR]{cross-stage partial super-resolution}
\newacro{csp}[CSP]{cross-stage partial}
\newacro{opdn}[OPDN]{omnidirectional position-aware deformable network}
\newacro{icip}[ICIP]{IEEE International Conference on Image Processing}

\begin{document}
\let\WriteBookmarks\relax
\def\floatpagepagefraction{1}
\def\textpagefraction{.001}

\shorttitle{360-Degree Video Super Resolution and Quality Enhancement Challenge: Methods and Results}
\shortauthors{Ahmed Telili, et~al.}

\title [mode = title]{360-Degree Video Super Resolution and Quality Enhancement Challenge: Methods and Results}                      
\sloppy
\author[1]{Ahmed Telili}[type=editor,
                        auid=,bioid=,
                        prefix=,
                        role=,
                        orcid=]

\author[1]{Wassim Hamidouche}[type=editor,
                        auid=,bioid=,
                        prefix=,
                        role=,
                        orcid=]

\author[1]{Ibrahim Farhat}[type=editor,
                        auid=,bioid=,
                        prefix=,
                        role=,
                        orcid=]
                        
\author[2]{Hadi Amirpour}[orcid=0000-0001-9853-1720]
\author[2]{Christian Timmerer}[orcid=0000-0001-9853-1720]

\author[1]{Ibrahim Khadraoui}[type=editor,
                        auid=,bioid=,
                        prefix=,
                        role=,
                        orcid=]

\author[3]{Jiajie Lu}[type=editor,
                        auid=,bioid=,
                        prefix=,
                        role=,
                        orcid=]

\author[4]{The Van Le}[type=editor,
                        auid=,bioid=,
                        prefix=,
                        role=,
                        orcid=]

\author[4]{Jeonneung Baek}[type=editor,
                        auid=,bioid=,
                        prefix=,
                        role=,
                        orcid=]

\author[4]{Jin Young Lee}[type=editor,
                        auid=,bioid=,
                        prefix=,
                        role=,
                        orcid=]

\author[2]{Yiying Wei}[type=editor,
                        auid=,bioid=,
                        prefix=,
                        role=,
                        orcid=]
\author[5]{Xiaopeng Sun}[type=editor,
                        auid=,bioid=,
                        prefix=,
                        role=,
                        orcid=]

\author[5]{Yu Gao}[type=editor,
                        auid=,bioid=,
                        prefix=,
                        role=,
                        orcid=]

\author[5]{JianCheng Huang}[type=editor,
                        auid=,bioid=,
                        prefix=,
                        role=,
                        orcid=]

\author[5]{Yujie Zhong}[type=editor,
                        auid=,bioid=,
                        prefix=,
                        role=,
                        orcid=]

\address[1]{Technology Innovation Institute, P.O.Box: 9639, Masdar City Abu Dhabi, UAE}                      
\address[2]{Christian Doppler Laboratory ATHENA, Alpen-Adria-Universität, Klagenfurt, Austria}

\address[3]{Politecnico di Milano, Milano, Italy}

\address[4]{Intelligent Visual Computing Lab. (IVCL), Sejong University, Seoul, South Korea}
\address[5]{Meituan Inc. China}

\begin{abstract}
Omnidirectional (360-degree) video is rapidly gaining popularity due to advancements in immersive technologies like \ac{vr} and \ac{xr}. However, real-time streaming of such videos, particularly in live mobile scenarios such as \acp{uav}, is hindered by limited bandwidth and strict latency constraints. While traditional methods such as compression and adaptive resolution are helpful, they often compromise video quality and introduce artifacts that diminish the viewer's experience. Additionally, the unique spherical geometry of 360-degree video, with its wide field of view, presents challenges not encountered in traditional 2D video. To address these challenges, we initiated the 360-degree Video Super Resolution and Quality Enhancement challenge. This competition encourages participants to develop efficient \ac{ml}-powered solutions to enhance the quality of low-bitrate compressed 360-degree videos, under two tracks focusing on $2\times$ and $4\times$ \ac{sr}. In this paper, we outline the challenge framework, detailing the two competition tracks and highlighting the \ac{sr} solutions proposed by the top-performing models. We assess these models within a unified framework, (\textit{i}) considering quality enhancement, (\textit{ii}) bitrate gain, and (\textit{iii}) computational efficiency. \ {Our findings show that lightweight single-frame models can effectively balance visual quality and runtime performance under constrained conditions, setting strong baselines for future research. These insights offer practical guidance for advancing real-time 360-degree video streaming, particularly in bandwidth-limited immersive applications.}

\end{abstract}

\maketitle

\section{Introduction}
In recent years, advancements in immersive video technologies, such as \ac{vr} and \ac{xr}, have revolutionized user experiences, enabling individuals to step into lifelike digital environments that closely mimic real-world scenarios. Several visual media formats, including \ac{odv}, light fields, and volumetric videos, are widely employed to create immersive viewing experiences~\cite{10268867}. In particular, \ac{odv}, also known as 360-degree video, has gained prominence due to the increasing availability of both capture and display technologies, as well as standardization efforts that facilitate interoperability~\cite{10582866}.

\Acp{odv}, although offering a broad field of view, suffer from reduced angular resolution due to their capture method using a fisheye lens on a sensor designed for planar images. This limitation can result in a loss of detail and clarity, potentially compromising the viewer's sense of immersion. Consequently, capturing and delivering high-resolution \acp{odv} is crucial for ensuring quality and user satisfaction in immersive environments. However, real-time streaming of high-resolution \acp{odv}, particularly in mobile scenarios like live \acp{uav} applications, presents significant challenges due to bandwidth and latency constraints. Although compression and adaptive resolution techniques strive to balance between quality and latency, the processes of downscaling and encoding can inevitably lead to quality degradation and the introduction of compression artifacts, especially in scenarios with severely limited bandwidth.

\begin{figure*}[pos=t]
\centering
\includegraphics[width=0.93\linewidth]{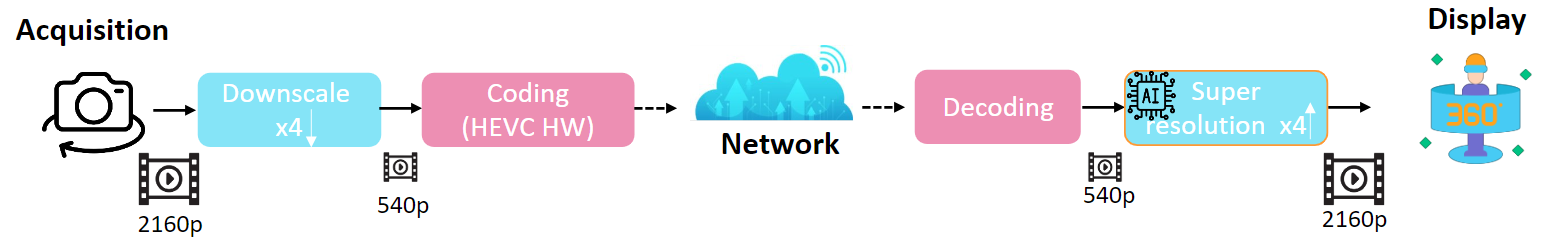}
\caption{Example for a $\times4$ super resolution integration in a typical streaming pipeline.}
\label{pipeline}
\end{figure*}

Deep learning-based methods have demonstrated remarkable success in enhancing the quality of image and video \ac{sr}. Leveraging techniques such as \ac{cnn}~\cite{dong2014learning,yamanaka2017fast,9136874,tian2020lightweight,9105085}, \ac{rnn}~\cite{9730760, 8425771}, \ac{gan}~\cite{8099502,9377002}, \ac{vit}~\cite{9607618, 10204527,10023973}, and diffusion models~\cite{9878449}, these methods have significantly advanced the state-of-the-art in image quality and detail enhancement.  While 360-degree images and videos are often converted into 2D planar representations like \ac{erp} or \ac{cmp} for practical purposes, these transformations introduce inherent distortions that are typically not addressed by conventional 2D \ac{sr} algorithms. Fig.~\ref{pipeline} depicts a typical \ac{odv} streaming pipeline. In bandwidth-constrained environments, it becomes crucial to consider both downscaling and compression distortions to optimize visual quality and, subsequently, improve the users' \ac{qoe}. Although recent research has proposed methods specifically designed for omnidirectional image and video \ac{sr}~\cite{10102571,9506233,ozcinar2019super,yu2023osrt}, most of these approaches focus on upscaling uncompressed content. There remains a scarcity of specialized techniques for enhancing compressed content, particularly at low bitrates. Moreover, the computational demands of these methods often prevent real-time video processing, even with the aid of modern hardware accelerators.

This paper presents the 360-Degree Video Super-Resolution and Quality Enhancement Challenge, proposed to address the critical challenges associated with low-bitrate streaming of \ac{odv}. The challenge aims to advance current research and development efforts by enhancing the real-time super-resolution capabilities of \ac{ml}-based algorithms, thereby improving the \ac{qoe} for end-users in live, low-bitrate streaming scenarios. Organized in conjunction with the 2024 \ac{icip}, the challenge features two tracks, targeting \ac{odv} resolution enhancement factors of  $2\times$ and  $4\times$. Further, this challenge is intended to evaluate the efficacy of \ac{sr} algorithms under realistic, bandwidth-constrained, and real-time conditions. With an initial registration of 45 participants across both tracks, the final testing stage received submissions from four teams, including their results, source codes, and fact sheets. {Distinctively, this challenge introduces several innovative aspects compared to previous works. First, we propose a novel unified evaluation framework that integrates a composite metric score balancing both visual quality, measured using \acs{ws-psnr}, and computational efficiency,  assessed by runtime, explicitly designed to reflect practical real-time application constraints. Second, we explicitly address the unique demands of low-bitrate compressed omnidirectional video streaming, thereby filling a critical gap in existing \ac{sr} challenges. Lastly, we introduced, ODVista, a new dataset designed to enable realistic benchmarking by incorporating multiple compression levels, scaling factors, and diverse scenes reflective of actual streaming scenarios encountered in applications such as those encountered in UAV-based applications.}

Table \ref{tab:challenges} summarizes the key characteristics of recent challenges focused on addressing the problem of image and video \ac{sr}~\cite{9857103, 10208813, 10208706, ignatov2022efficient, ignatov2022power, yang2022aim}. In contrast to these existing challenges, ours uniquely considers real-time constraints and compression distortions for \ac{odv} content. Specifically, both quality enhancement (measured in \acs{ws-psnr}) and runtime performance are integrated into the final scoring metric used to rank submitted solutions.

\begin{table*}[]
\centering
\caption{Recent challenges in image and video \ac{sr}. The performance metrics used to rank the models are highlighted in bold. }

\adjustbox{width=\textwidth}{
\label{tab:challenges}
\begin{tabular}{@{}
>{\columncolor[HTML]{FFFFFF}}l 
>{\columncolor[HTML]{FFFFFF}}c 
>{\columncolor[HTML]{FFFFFF}}c 
>{\columncolor[HTML]{FFFFFF}}c 
>{\columncolor[HTML]{FFFFFF}}c 
>{\columncolor[HTML]{FFFFFF}}c 
>{\columncolor[HTML]{FFFFFF}}c 
>{\columncolor[HTML]{FFFFFF}}c 
>{\columncolor[HTML]{FFFFFF}}c 
>{\columncolor[HTML]{FFFFFF}}c 
>{\columncolor[HTML]{FFFFFF}}c @{}}
\toprule
Challenge (year) & Platform & Image  & Video & Compression & 360-degree & \acs{sr} ratio  & Quantisation & Quality & Efficiency   & Dataset  \\ \midrule
NITRE workshop (2022)~\cite{9857103} & \xmark & \xmark & \cmark & \cmark & \xmark &  1$\times$, 2$\times$, 4$\times$  & \xmark &  {\bf \acs{psnr}} & \xmark  &  LDV 2.0  \\  \midrule
NITRE workshop (2023)~\cite{10208813} & RTX3090 GPU & \cmark &  \xmark & \xmark & \xmark & 2$\times$, 3$\times$  & \xmark & {\bf \acs{psnr}}, \acs{ssim} & \begin{tabular}[c]{@{}c@{}} {\bf Runtime}, \acs{flop}, \\ \#params \end{tabular}   & 4K RTSR   \\  \midrule
NITRE workshop (2023)~\cite{10208706}  & \xmark & \cmark & \cmark & \xmark & \cmark  & 4$\times$ &   \xmark & {\bf \acs{ws-psnr}} & \xmark  &  Flickr360 \& ODV360 \\ \midrule
 AI \& AIM
workshop (2022)~\cite{ignatov2022efficient} &  \acs{npu} & \cmark & \xmark & \xmark & \xmark & 3$\times$ &   \begin{tabular}[c]{@{}c@{}} INT8, FP16, \\  
 or FP32 \end{tabular} & {\bf \acs{psnr}}, \acs{ssim} & {\bf Runtime}, model size & DIV2K  \\  \midrule
AI \& AIM
workshop (2022)~\cite{10.1007/978-3-031-25066-8_6} & \acs{npu} & \xmark & \cmark & \xmark & \xmark & 4$\times$ &  \begin{tabular}[c]{@{}c@{}} INT8, FP16, \\  
 or FP32 \end{tabular}  & {\bf\acs{psnr}}, \acs{ssim} & \begin{tabular}[c]{@{}c@{}} Runtime, {\bf power} \\ model size \end{tabular}  &  REDS~\cite{9025509}  \\  \midrule
AI \& AIM
workshop (2022)~\cite{ignatov2022efficient} & GPU & \cmark & \cmark & \cmark & \xmark & 4$\times$  & \xmark & {\bf \acs{psnr}} &  Runtime &  DIV2K \& LDV 3.0  \\  \midrule
Ours
ICIP challenges (2024) & RTX6000 GPU  & \xmark & \cmark & \cmark & \cmark & 2$\times$, 4$\times$ &  \xmark & \begin{tabular}[c]{@{}c@{}} {\bf \acs{ws-psnr}}, \\  
\acs{ws-ssim}  \end{tabular} & \begin{tabular}[c]{@{}c@{}} {\bf Runtime}, \acs{flop}, \\  
\#params \end{tabular}  & ODVista~\cite{telili2024odvista}   \\
\bottomrule
\end{tabular}
}
\end{table*}

The remainder of this paper is organized as follows. Section~\ref{sec:related_work} provides a comprehensive review of existing methods for 360-degree image and video super-resolution. Section~\ref{sec:challenge} introduces the challenge and elaborates on its setup, including details on the dataset utilized, the challenge framework, and the specific tracks involved. Section~\ref{sec:methods} offers a detailed description of the various approaches and techniques proposed by the participants, while Section~\ref{sec:results} presents and analyzes the experimental results. Finally, Section~\ref{sec:conclusion} concludes this paper.

\section{Related Work}
\label{sec:related_work}
This section reviews existing \ac{sr} methods, encompassing both conventional 2D and 360-degree approaches, categorized by their input modality namely, image (Section \ref{sec:rw_imagesr}) and video (Section \ref{sec:rw_videosr}), and their computational efficiency (Section \ref{sec:rw_rtperformance}).

\subsection{Omnidirectionnal image-super resolution}
\label{sec:rw_imagesr}
Dong~\etal~\cite{7115171} pioneered the use of \acp{dnn} for \ac{sisr}. Subsequently, \ac{sisr} performance has significantly improved by leveraging more advanced deep learning architectures, such as \acp{cnn}~\cite{8954252, 7780551, 10.1007/978-3-319-46475-6_25, 9857155, 8014885, 9578003, 10.1007/978-3-030-58610-2_12}, \acp{gan}~\cite{8099502, 9607421, 10.1007/978-3-030-11021-5_5, 9497054}, and more recently, \acp{vit}~\cite{9577359, 10023973, 10204527, li2022efficient, 9607618, 10377177} and diffusion models~\cite{9878449}. \Ac{srcnn}~\cite{7115171} exemplifies the use of a \ac{cnn}-based architecture trained end-to-end to learn the mapping between \ac{lr} and \ac{hr} image representations. Lim~\etal~\cite{8014885} effectively utilize \ac{resnet} for image \ac{sr}, yielding quality improvements. SwinIR~\cite{9607618}, a robust baseline model for image \ac{sr}, is built upon the Swin Transformer architecture~\cite{9710580}. It comprises multiple \acp{rstb}, each containing several Swin transformer layers interconnected with residual connections. Evaluated on various downstream tasks, including image SR, image denoising, and JPEG compression artifact reduction, SwinIR demonstrates superior performance compared to state-of-the-art models. It achieves a notable \ac{psnr} improvement of \qty{0.14} to \qty{0.45}{\decibel} while simultaneously reducing the number of parameters by up to \qty{67}{\percent}. In other research, adversarial training with synthetic data~\cite{9710356} has incorporated various types of image degradation to further enhance the performance of \ac{sisr} models. More recently, Chen~\etal~\cite{10204527} proposed a \ac{hat} that combines channel attention with window-based self-attention mechanisms, leveraging the former's ability to utilize global statistics and the latter's strong local fitting capability. Additionally, they introduce an overlapping cross-attention module to enhance interaction between neighboring window features. The resulting model, pre-trained on the same task, outperforms state-of-the-art methods by more than \qty{1}{\decibel} (\acs{psnr}). \Ac{dat}~\cite{10377177} is another Transformer-based \ac{sisr} model that aggregates features across spatial and channel dimensions in a dual inter- and intra-block manner. Specifically, spatial and channel self-attention in consecutive transformer blocks are used, enabling the \ac{dat} model to capture the global context and realize inter-block feature aggregation. Furthermore, they propose the \ac{aim} and the \ac{sgfn} to achieve intra-block feature aggregation.

Numerous studies have investigated omnidirectional image \ac{sr} in projection formats. Notably, Fakour-Sevom~\etal~\cite{visapp18} train the \ac{srcnn} model on omnidirectional images in the \ac{erp} format to learn how to recover \ac{hr} images and mitigate projection artifacts. The resulting model demonstrates a significant improvement in super-resolved image quality, achieving an average \ac{psnr} gain of \qty{1.36}{\decibel} compared to the baseline bicubic interpolation method. The perceptual quality of this model is further enhanced by incorporating \ac{gan} training~\cite{8901764, 9150746}. \Acp{odi} in~\ac{erp} format exhibit non-uniform pixel density and texture complexity across different latitudes. The \ac{lau-net}~\cite{9577746} addresses this characteristic in \ac{odi} \ac{sr} by employing a Laplacian multi-level separation architecture to divide the \ac{odi} into distinct latitude bands. Subsequently, a deep reinforcement learning scheme with latitude-adaptive reward is utilized to select the optimal upscaling factor for each individual band. Nishiyama~\etal~\cite{9506233} address this issue by incorporating the distortion map as an additional input to the \ac{sisr} network and augmenting the training dataset with artificially distorted perspective images. SphereSR~\cite{9880040} learns \ac{sr} directly in the continuous spherical domain by extracting features on the spherical surface and utilizing a \ac{sliif} to predict RGB values from spherical coordinates. This approach enables the generation of HR images from LR 360-degree input in any desired projection format. Finally, OSRT~\cite{10203515} is a distortion-aware Transformer-based solution that continuously and self-adaptively modulates \ac{erp} distortions by learning deformable offsets directly from \ac{erp} distortion maps.

The winning model~\cite{10208527} in the NTIRE 360-degree image super-resolution challenge~\cite{10208706} employs a two-stage architecture. The first stage introduces a novel \ac{opdb}, effectively integrating dimensional information and absolute position encoding within 360-degree images. The second stage, sharing the same architecture but incorporating pixel unshuffling and downsampling operations, is fine-tuned using the weights learned in the first stage.
The second-ranking model utilizes an architecture that synergistically combines Transformers with \ac{csa} and \ac{wt}. It undergoes a two-phase training process, where the spatial resolution of patches is progressively increased in the second phase, employing the L1 norm loss function for optimization. The third-ranking model employs an architecture that first extracts local structural information through shift convolutions, maintaining the same level of complexity as standard 1$\times$1 convolutions. Additionally, the model incorporates group-wise \ac{opdb} for self-attention computation on non-overlapping groups of features, utilizing various window sizes to capture long-range dependencies effectively. These three top-performing models demonstrated substantial improvements in \ac{ws-psnr}, achieving gains of \qty{0.78}{\decibel}, \qty{0.63}{\decibel}, and \qty{0.42}{\decibel}, respectively, compared to the SwinIR baseline~\cite{10208706}. However, as inference efficiency was not a consideration in this challenge, the top models exhibit high complexity and a parameter count comparable to SwinIR.

\subsection{Omnidirectionnal video super-resolution}
\label{sec:rw_videosr}
\Ac{vsr} leverages the inherent temporal dependency among neighboring frames to enhance the recovery of fine details and maintain temporal consistency. Several solutions have been proposed to address \ac{vsr} challenges, primarily relying on either the sliding-window framework, where each frame is restored using a short temporal window, or the recurrent framework, which exploits long-term dependencies through latent feature propagation. However, the latter approach introduces additional design complexities, involving the propagation, alignment, concatenation, and upsampling of information, leading to \ac{vsr} models that are inherently more intricate than their \ac{sisr} counterparts. A selection of notable \ac{vsr} models is reviewed in the subsequent paragraphs.

TDAN~\cite{9156615} utilizes \acp{dcn} to align the features of adjacent frames. EDVR~\cite{9025464} adopts multiple attention layers and a multi-scale \acp{dcn} alignment module for aligning and concatenating features from different frames. FRVSR~\cite{frvsr} relies on an end-to-end trainable frame recurrent video \ac{sr} framework that uses the previously inferred \ac{hr} estimate to super-resolve the subsequent frame. This approach produces compelling and temporally consistent output while reducing computational costs by warping only one image per step. MuCAN~\cite{10.1007/978-3-030-58607-2_20} presents an effective multi-correspondence aggregation network for \ac{vsr}. The model relies on a temporal multi-correspondence aggregation strategy to leverage similar patches across frames, and a cross-scale nonlocal-correspondence aggregation scheme to explore self-similarity of images across scales. BRCN~\cite{7919264} proposes a bidirectional recurrent convolutional network for efficient \ac{vsr}. It replaces the commonly used recurrent full connections in \acp{rnn} with weight-sharing convolutional connections and adds conditional convolutional connections from previous input layers to the current hidden layer for enhancing visual-temporal dependency modeling. RSDN~\cite{10.1007/978-3-030-58610-2_38} addresses the challenges of changing appearance and error accumulation through a recurrent detail structural block and a hidden state adaptation module. BasicVSR~\cite{9577681} offers a simple yet effective model that relies on a bidirectional propagation scheme to maximize information gathering, coupled with an optical flow-based method for precise feature alignment between neighboring frames. Utilizing straightforward concatenation and pixel-shuffle for aggregation and upsampling, BasicVSR surpasses existing solutions in both quality enhancement and computational efficiency.


While general \ac{vsr} models are not explicitly designed for 360-degree videos, several studies have addressed the unique distortions inherent to omnidirectional video content~\cite{9155477, LIU2024108601, 10102571}. The top-performing \ac{vsr} models submitted to the 360-degree video challenge~\cite{10208706}, ranked based on their \ac{ws-psnr}, achieved notable gains over the BasicVSR baseline model on the test set. These improvements amounted to \qty{0.47}{\decibel}, \qty{0.84}{\decibel}, and \qty{1.32}{\decibel}, respectively. The best-performing model utilizes BasicVSR++~\cite{9879156}, an enhanced version of BasicVSR. This improvement incorporates second-order grid propagation and flow-guided deformable alignment, facilitating superior information propagation and aggregation within the \ac{vsr} process. Furthermore, this solution employs a multi-stage training strategy with four distinct phases to optimize performance. The second-best model integrates an \ac{erp}-Adapter module into the BasicVSR++ framework to address \ac{erp} distortions. This module effectively transfers knowledge from a pre-trained perspective \ac{vsr} network to adapt to \ac{erp} videos with minimal additional training. The \ac{erp}-Adapter leverages deformable convolutions to modulate \ac{erp} features and an adaptive spatial transformer network (AdaSTN). The AdaSTN indirectly calculates offsets by estimating the pixel-level affine transformation matrix and translation vector. This model's training occurs in two steps: first, BasicVSR++ is trained, then the adapter is fine-tuned while freezing BasicVSR++. The third-ranked solution employs a two-stage model for \ac{vsr}. The first stage performs 4$\times$ \ac{sisr} using the \ac{hat}-L model~\cite{10204527}. The resulting high-resolution samples are then downsampled using PixelUnshuffle and processed with a 1$\times$1 convolution layer to adjust the feature dimension before being fed into a BasicVSR++ model. This multi-stage approach effectively bridges the gap between \ac{sisr} and \ac{vsr}, leveraging the strengths of both techniques to achieve the desired enhanced resolution.

More recently, Baniya~\etal~\cite{10102571} proposed Spherical Signal \ac{sr} with Proportioned Optimization (S3PO) to specifically address the distortions in 360-degree video signals. S3PO replaces the traditional alignment mechanism with a sequential modeling approach within a recurrent architecture. This allows for the extraction of relevant features directly from unaligned frames without explicit alignment, leveraging an attention mechanism that adapts focus across both spatial and temporal dimensions. The extracted local and global features are then fused and processed through a series of dual-duct residual blocks to reconstruct the high-resolution output. The quality achieved by S3PO is marginally superior to that of the BasicVSR model.

\subsection{Computational efficiency for real-time processing}
\label{sec:rw_rtperformance}
The application of \acp{dnn} to image and video \ac{sr} has led to remarkable advancements in quality enhancement compared to traditional hand-crafted approaches. However, despite leveraging parallel processing capabilities of modern \ac{gpu} platforms, deep learning-based models often struggle to achieve real-time processing speeds, thus hindering their deployment in applications demanding stringent real-time performance.

\Ac{fsrcnn}~\cite{10.1007/978-3-319-46475-6_25} aims to accelerate \ac{srcnn} by implementing three key modifications: ({\it i}) replacing the initial bicubic interpolation step with a deconvolution layer at the network's end, ({\it ii}) learning the mapping directly from \ac{lr} image features instead of \ac{hr} ones using a reformulated mapping layer with reduced input dimensions, and ({\it iii}) employing smaller convolutional kernels across a deeper architecture with more mapping layers. \Ac{fsrcnn} achieves a speedup exceeding 40 times that of \ac{srcnn} while also demonstrating superior image quality enhancement.

The NTIRE 2023 Real-Time 4K SR Challenge~\cite{10208813} established a framework for developing efficient real-time \ac{sisr} models capable of upscaling diverse content, such as digital art and photography, to ultra-high 4K resolution. The ranking of the solutions is determined by a composite score that incorporates both image quality and computational efficiency, utilizing \ac{psnr} and runtime, respectively.  The top-performing solutions in this challenge leverage three key architectural design and training strategies: ({\it i}) re-parameterization for model compression and inference acceleration, ({\it ii}) pixel shuffling (unshuffling) for efficient upsampling (downsampling), and ({\it iii}) multi-stage training for progressive refinement of image details by alternating different learning rates and loss functions. Zamfir~\etal~\cite{10208555} propose a fast and efficient \ac{sr} model for upscaling 4K images from lower resolutions (e.g., 720p, 1080p). The authors employ a standard \ac{sisr} architecture, comprising shallow and deep feature extraction modules followed by an upsampling module. To reduce computational complexity, deep features are downscaled while simultaneously extracting high-frequency information from the \ac{lr} image. Furthermore, the NAFNet block~\cite{10.1007/978-3-031-20071-7_2} is utilized for efficient deep feature computation. Finally, the concept of re-parameterization~\cite{9577516} is applied during inference to further reduce the overall runtime. In addition, the depth-wise convolution in the attention mechanism is used to increase the receptive field along with a novel pixel normalization to improve training stability. To further enhance the \ac{sisr} tradeoff between efficiency and quality enhancement, ECBSR~\cite{10.1145/3474085.3475291} introduces a convolutional block that utilizes structural re-parameterization to enhance learning capability without sacrificing model efficiency.  Zhisheng~\etal~\cite{9857219} propose ESRT, which combines shallow \acp{cnn} for adaptive feature map size adjustment with a lightweight transformer. This architecture enables low-complexity feature extraction and captures long-term dependencies among patches with similar characteristics.

The Mobile AI challenges in 2022 for \ac{sisr}~\cite{ignatov2022efficient} and \ac{vsr}~\cite{10.1007/978-3-031-25066-8_6} focused on the impact of quantization and model design for edge and mobile devices. Solutions proposed for these challenges primarily employed ({\it i}) lightweight convolutional layers with small kernel sizes, ({\it ii}) re-parameterization techniques, and ({\it iii}) quantization techniques to reduce model size and improve efficiency, albeit with a slight trade-off in quality.

\begin{figure}
\centering
 \includegraphics[width=0.99\linewidth]{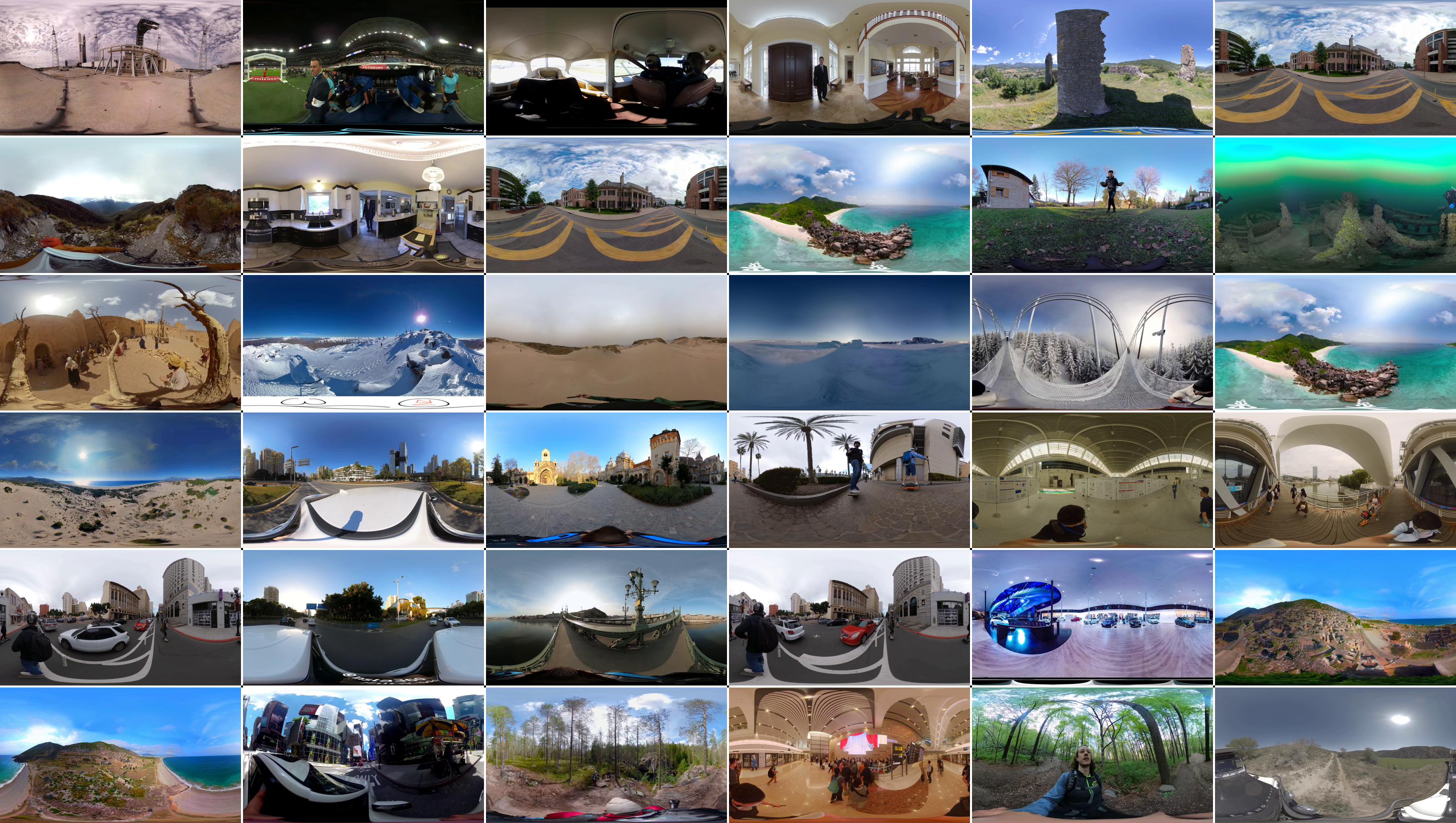}
\caption{Sample frames of the ODVista dataset.}
\label{frames}
\end{figure}

\section{360-Degree Video Super Resolution and Quality Enhancement Challenge}
\label{sec:challenge}
This challenge has three primary objectives: ({\it i}) to stimulate research in the development of real-time \ac{sr} techniques for \ac{odv} content, ({\it ii}) to provide a standardized platform for comparing diverse \ac{sr} methods under consistent evaluation conditions, particularly accounting for compression distortions, and ({\it iii}) to foster a collaborative environment that promotes the exchange of knowledge and innovation between academia and industry. The subsequent sections delve into the specifics of this challenge, outlining the dataset (Section \ref{sec:dataset}), tracks (Section \ref{sec:tracks}), and phases (Section \ref{sec:phases}).

\subsection{ODVista dataset}
\label{sec:dataset}
{The ODVista dataset~\cite{telili2024odvista} is specifically designed to advance research in omnidirectional or 360-degree \ac{vsr} and quality enhancement. It comprises 200 high-resolution videos, each encoded with the \ac{hevc}/H.265 standard at four distinct bitrate levels: 0.25,Mbps, 0.5,Mbps, 1,Mbps, and 2,Mbps. These videos are sourced from high-quality omnidirectional content, offering diverse visual complexity with a broad range of indoor and outdoor scenes to challenge \ac{sr} algorithms under varying conditions of texture, motion, and lighting. The videos are equally divided between 2K (1080p) and 4K (2160p) resolutions, with 100 videos in each category. They are collected from publicly available, high-quality sources including YouTube~\cite{youtube} and the ODV360 dataset~\cite{cao2023ntire}. Fig.~\ref{frames} shows sample frames of the \acs{vista} dataset.}

\begin{figure*}[pos=t]
\centering
\scriptsize
\centering
\begin{minipage}[b]{0.33\linewidth}
\centering
\centerline{\includegraphics[width=1\linewidth]{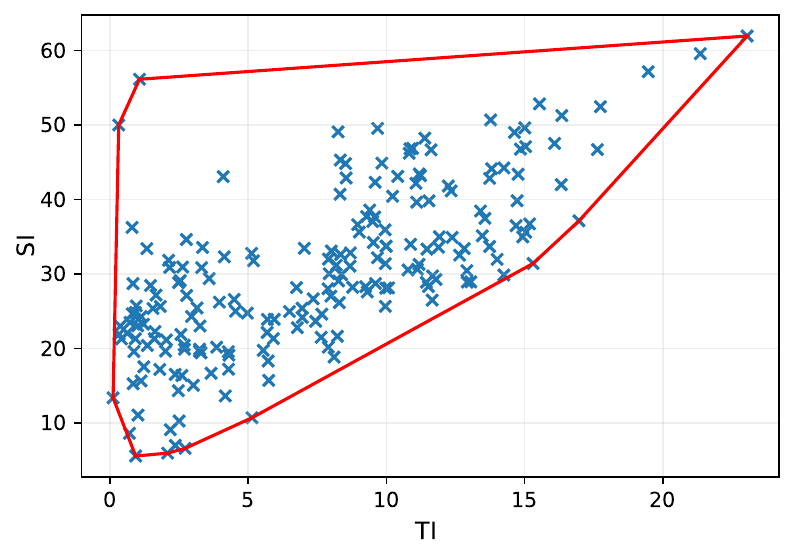}}
(a) SI vs. TI
\end{minipage}
\begin{minipage}[b]{0.33\linewidth}
\centering
\centerline{\includegraphics[width=1\linewidth]{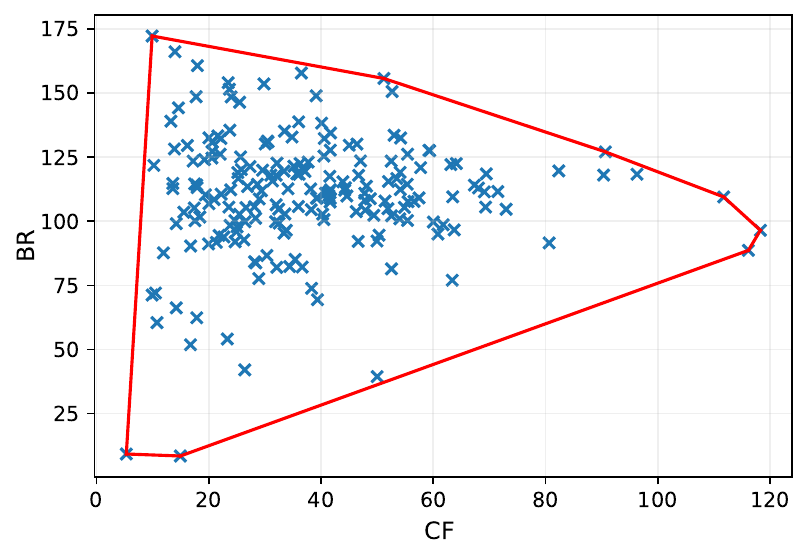}}
(b) BR vs. CF
\end{minipage}
\begin{minipage}[b]{0.33\linewidth}
\centering
\centerline{\includegraphics[width=1\linewidth]{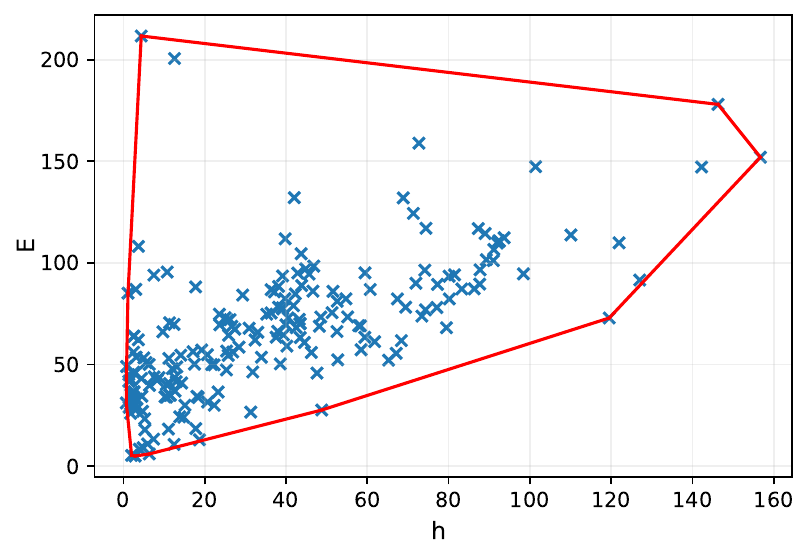}}
(b) h vs. E
\end{minipage}
\caption{{Source content distribution in paired feature space with corresponding convex hulls. Left column: TI versus SI, middle column: CF versus BR and right column: h versus E.}}
\label{features_plots}
\end{figure*}

{To further assess and validate the dataset’s diversity, we analyzed six low-level video descriptors: spatial information (SI), temporal information (TI), brightness range (BR), colorfulness (CF), and spatial/temporal complexity ($E$ and $h$) derived from \ac{vca}\cite{menon2022vca}. These features were averaged per video to produce a global representation. As illustrated in Fig.\ref{features_plots}, the dataset spans a broad spectrum of spatiotemporal complexity—ranging from low-motion, low-texture sequences to fast-changing and detail-rich content, and naturally includes various real-world distortions such as noise and motion blur. SI and TI values range from 5 to 62 and 0 to 24, respectively, while the BR–CF distribution reveals the presence of varied lighting environments and color palettes. This confirms that the ODVista dataset offers a well-balanced mix of content types, making it suitable for benchmarking \ac{vsr} models under realistic, bandwidth-constrained conditions.}

{To replicate authentic streaming scenarios, such as those involving \acp{uav}, each video is downscaled by factors of 2$\times$ and 4$\times$ and encoded using Nvidia’s NVENC hardware encoder~\cite{nvidia} to support efficient, real-time compression.  This process generates four distinct streams for each source video at the specified bitrates, resulting in a total of 2200 videos, including 1600 compressed and 600 uncompressed. The ODVista dataset further ensures a balanced representation of spatial and temporal complexities. To maintain consistent scene content within each clip, the videos are temporally segmented into 100-frame clips. The videos are stored in the equirectangular projection format, a standard for 360-degree videos, preserving their immersive quality and integrity.}

\subsection{Tracks}
\label{sec:tracks}
The proposed challenge encompasses two distinct tracks, targeting \ac{sr} ratios of $2\times$ and $4\times$, respectively.
\subsubsection{Track 1: 360-degree Video Super Resolution and Quality Enhancement $2\times $}
This track focuses on enhancing 360-degree videos that have undergone both compression and a 2$\times$ downscaling. Utilizing the ODVista dataset, participants are challenged to develop \ac{sr} models that effectively balance objective quality metrics with computational efficiency. The objective is to achieve a 2$\times$ upscale in resolution, with performance evaluated using a predefined scoring mechanism that considers both the quality of the enhancement and the runtime efficiency of the model.
\subsubsection{Track 2:  360-degree Video Super Resolution and Quality Enhancement $4\times$}
Analogous to Track 1 but with increased complexity, this track centers on videos that have undergone both compression and a 4$\times$ downscaling. Participants are tasked with developing \ac{sr} models, utilizing the ODVista dataset, to achieve the optimal balance between enhanced quality and computational efficiency. The objective is to achieve a 4 $\times$ resolution upscale, with performance evaluated using the same scoring mechanism as in Track 1, ensuring consistency in the assessment process.

\subsection{Framework} {To streamline participant involvement and focus their efforts solely on model development, we have created an intuitive, modular framework available on GitHub\footnote{\url{https://github.com/Omnidirectional-video-group/360_VSR}, last access: April 22, 2025.}. This framework encompasses pre-built modules for data handling, training procedures and evaluation metrics. Specifically, it includes clearly defined scripts and configurations for dataset loading, preprocessing, and automatic evaluation using our defined metrics (\acs{ws-psnr}, runtime performance). Participants only need to integrate their novel \ac{sr} model architectures into the provided model module, simplifying experimentation and promoting reproducibility.}

\subsection{Phases}
\label{sec:phases}
\vspace{1em}
\subsubsection{Development and validation phase} 
Participants were provided with the training data and low-resolution encoded versions of the validation set from the ODVISTA dataset. While the ground truth high-resolution validation videos were withheld, participants had the opportunity to train their models on the provided data or additional external data and upload their resulting high-resolution validation outputs to the evaluation server for performance assessment. Live score values were subsequently uploaded to the CodaLab platform\footnote{\url{https://codalab.lisn.upsaclay.fr/competitions/17458}, last access: July 31, 2024.}, facilitating regular updates to the leaderboard showcasing team rankings.

\subsubsection{Testing phase} 
The full validation dataset was subsequently made available at the testing phase, enabling participants to further refine their models with a more extensive and diverse range of data. Upon the conclusion of this phase, participants were required to submit their trained models accompanied by their corresponding source code. To ensure a fair and consistent evaluation, we then proceeded to evaluate their models locally on the same designated device, utilizing the ODVISTA test set.

\section{Challenge Methods and Teams}
\label{sec:methods}
Out of the 47 initial participants registered to the challenge, 10 actively engaged in the evaluation phase. Ultimately, four teams successfully progressed to the final phase, submitting valid results for consideration. This section provides a comprehensive overview of the diverse methodologies employed by each of these 4 participating teams. Among these finalists, Team FFCIR was represented by Jiajie Lu from Politecnico di Milano, Italy. Team IVCL from the Intelligent Visual Computing Lab, Sejong University, South Korea, included The Van Le, Jeonneung Baek, and Jin Young Lee. Team VACV from Meituan Inc., China, consisted of Xiaopeng Sun, Yu Gao, JianCheng Huang, and Yujie Zhong. Team ATHENA was represented by Yiying Wei from Christian Doppler Laboratory ATHENA at Alpen-Adria-Universität, Klagenfurt, Austria.



\subsection{FFCIR Team}
\begin{figure*}[pos=t]
    \centering
    \includegraphics[width=0.8\textwidth]{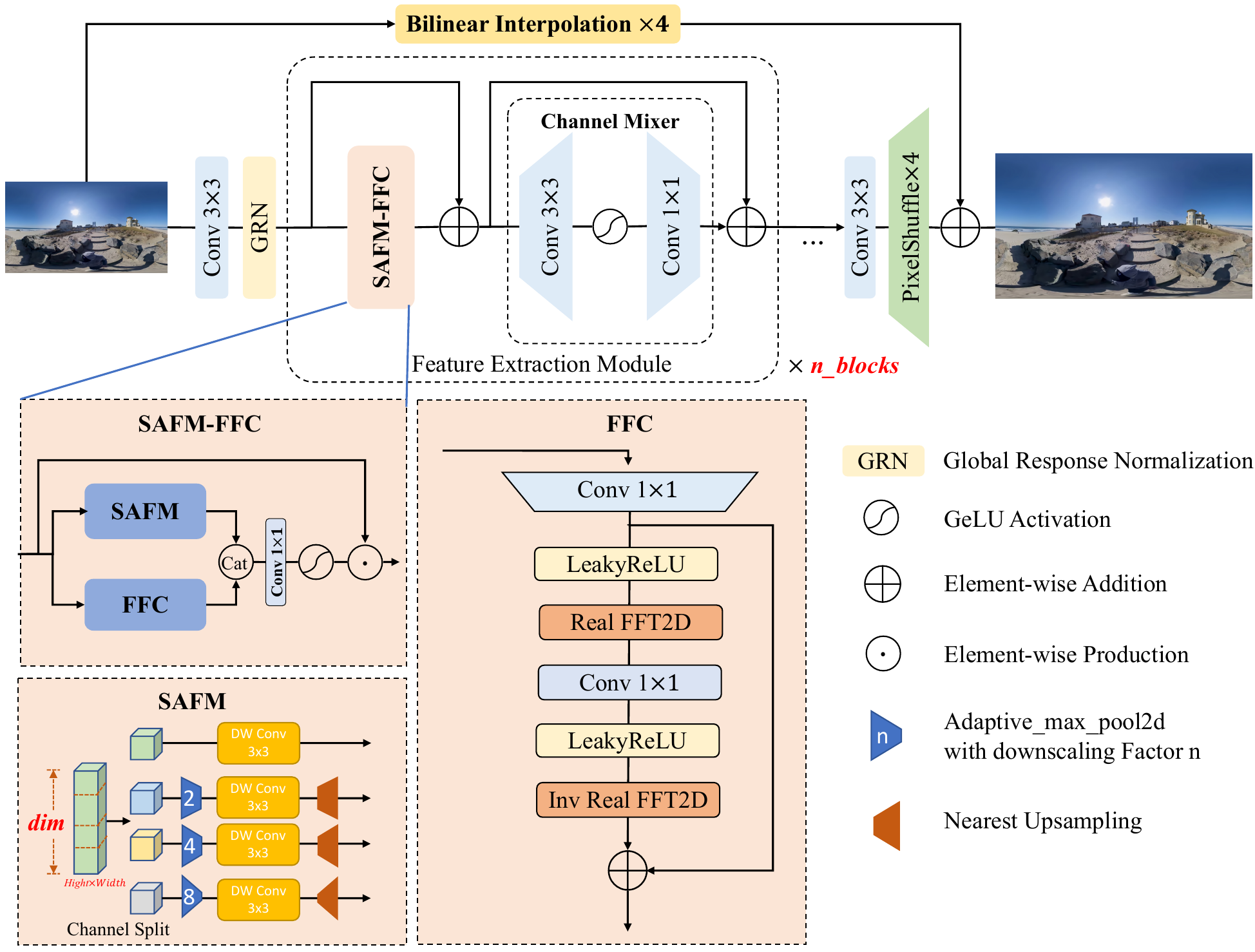}
    \caption{Team FFCIR: Frequency-Domain Enhanced Spatially-Adaptive Feature Modulation for super resolution.}
    \label{fig:FFCIR}
\end{figure*}

\vspace{1em}
\noindent \textbf{Network Architecture.}
The network architecture adopts the design of SAFMN~\cite{SAFMN_2023_ICCV}, as illustrated in Fig.~\ref{fig:FFCIR}. However, it removes the LayerNorm from the original method to reduce runtime and incorporates a  \ac{grn}~\cite{woo2023convnext} after the initial 3$\times$3 convolution. \Ac{safm} employs a feature pyramid strategy to uncover long-range information in feature maps at various scales. Subsequently, the reassembled features are processed through a Channel Mixer to extract local information. \\\\
\textbf{Frequency-Domain Enhancement.}
Drawing inspiration from \ac{ffc}~\cite{chi2020fast} and SwinFIR~\cite{swinfir}, an \ac{ffc} block parallel to \ac{safm} is integrated into the deep image feature extraction process to capture frequency-domain information. All features are then concatenated to aggregate both spatial and frequency domain features for further processing. This design effectively improves the recovery of high-frequency regions in the image at a minimal computational cost. \\\\
\textbf{Training Strategy.}
The number of blocks (\textit{n\_blocks}) in Fig.~\ref{fig:FFCIR} and the number of channels were set to \num{8} and \num{36}, respectively. The model was trained from scratch using 128$\times$128 patches randomly cropped from \ac{lr} images. Optimization was achieved using a combination of Charbonnier loss~\cite{charbonnier1994two} and an \ac{fft} loss~\cite{Yadav_2021}, minimized with the Adam optimizer. Data augmentation techniques, including flipping, rotation, and mixup, were employed. The initial learning rate was set to \num{2d-4}, with a total of \num{8d5} iterations and a batch size of \num{16}.

\subsection{IVCL Team}

\begin{figure*}[pos=t]
    \centering
    \includegraphics[width=0.8\linewidth]{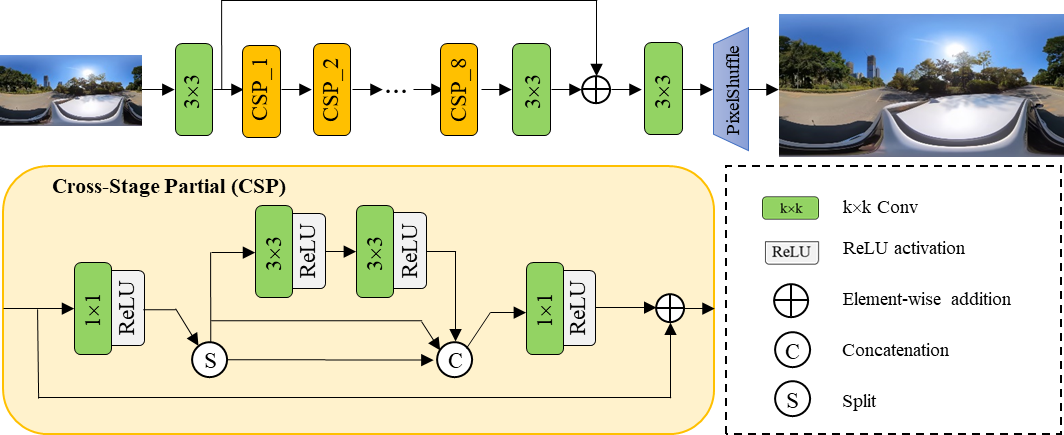}
    \caption{Team IVCL: Cross-Stage Partial Super Resolution (CSPSR).}
    \label{fig:Panel}
\end{figure*}

\vspace{1em}
\noindent \textbf{Network Architecture.} The IVCL team proposed a \ac{cspsr} network for efficient 360-degree video super-resolution. As illustrated in Fig.~\ref{fig:Panel}, the network comprises three components: ({\it i}) shallow feature extraction, ({\it ii}) deep feature extraction, and ({\it iii}) reconstruction modules. In the shallow feature extraction module, low-frequency features are generated using a 3$\times$3 convolutional layer. High-frequency features are then extracted via \ac{csp} blocks in the deep feature extraction module. The \ac{csp} blocks, designed for rapid feature extraction in SR, are based on a CSPNet architecture \cite{wang2020cspnet}. For fast reconstruction, a 3$\times$3 Conv layer followed by a pixel-shuffle layer is employed to generate the 4$\times$ high-resolution output. \\\\
\textbf{Training Strategy.}
The \ac{cspsr}, comprising \num{8} \ac{csp} blocks, was trained with an input patch size of 64$\times$64. Data augmentation techniques, including random cropping, flipping, and rotation, were employed. The total number of iterations and batch size were set to \num{1d6} and \num{32}, respectively. The Charbonnier loss~\cite{charbonnier1994two} was utilized as the loss function and minimized using the Adam optimizer. The learning rate was initialized at \num{1d-4}.

\subsection{VACV Team}

\begin{figure*}[pos=t]
    \centering
    \includegraphics[width=0.8\textwidth]{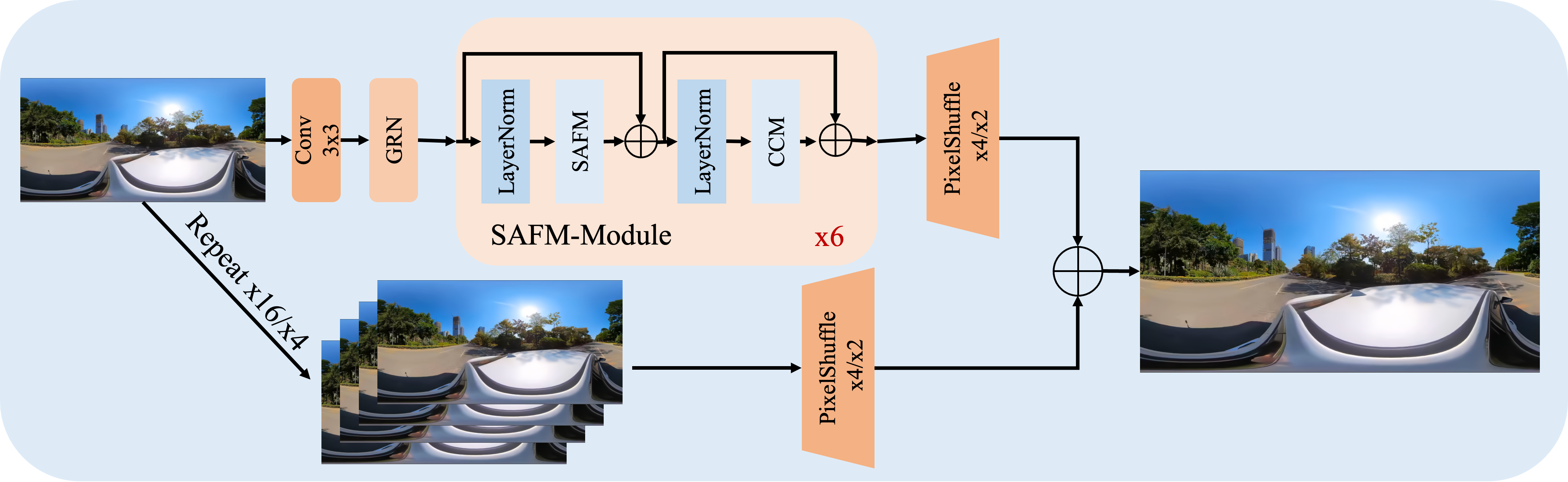}
    \caption{Team VACV: Efficient 360-degree super resolution via feature repetition.}
    \label{fig:vacv}
\end{figure*}

\vspace{1em}
\noindent \textbf{Network Architecture.} 
The VACV team proposed an efficient 360-Degree \ac{sr} approach leveraging feature repetition for efficient 360-degree \ac{vsr}. As illustrated in Fig.~\ref{fig:vacv}, the network consists of three parts: ( {\it i}) shallow feature extraction, ({\it ii}) deep feature extraction, and ({\it iii}) reconstruction module. In the shallow feature extraction phase, low-frequency features are generated through a 3$\times$3 convolutional layer. In the deep feature extraction phase, high-frequency features are extracted using the SAMFN module~\cite{SAFMN_2023_ICCV}. To expedite the reconstruction process, the low-resolution features were duplicated by the scaling factor. This technique not only enhances the model's performance but also accelerates inference speed. \\\\
{\bf Training Strategy.}
The model was trained with an input patch size of 64$\times$64. Data augmentation was performed through random cropping, flipping, and rotation. The total number of iterations was \num{3d5} with a batch size of \num{9}. The loss function employed was the weighted L1 loss~\cite{sun2017weighted}, minimized through the Adam optimizer. The learning rate was set to \num{2d-4}. The \ac{opdn} module~\cite{sun2023opdn} was also experimented with but found to be ineffective, potentially due to the limited model size and its restricted capacity to learn spatial location information.

\subsection{ATHENA Team}

\begin{figure*}[pos=t]
    \centering
    \includegraphics[width=0.85\linewidth]{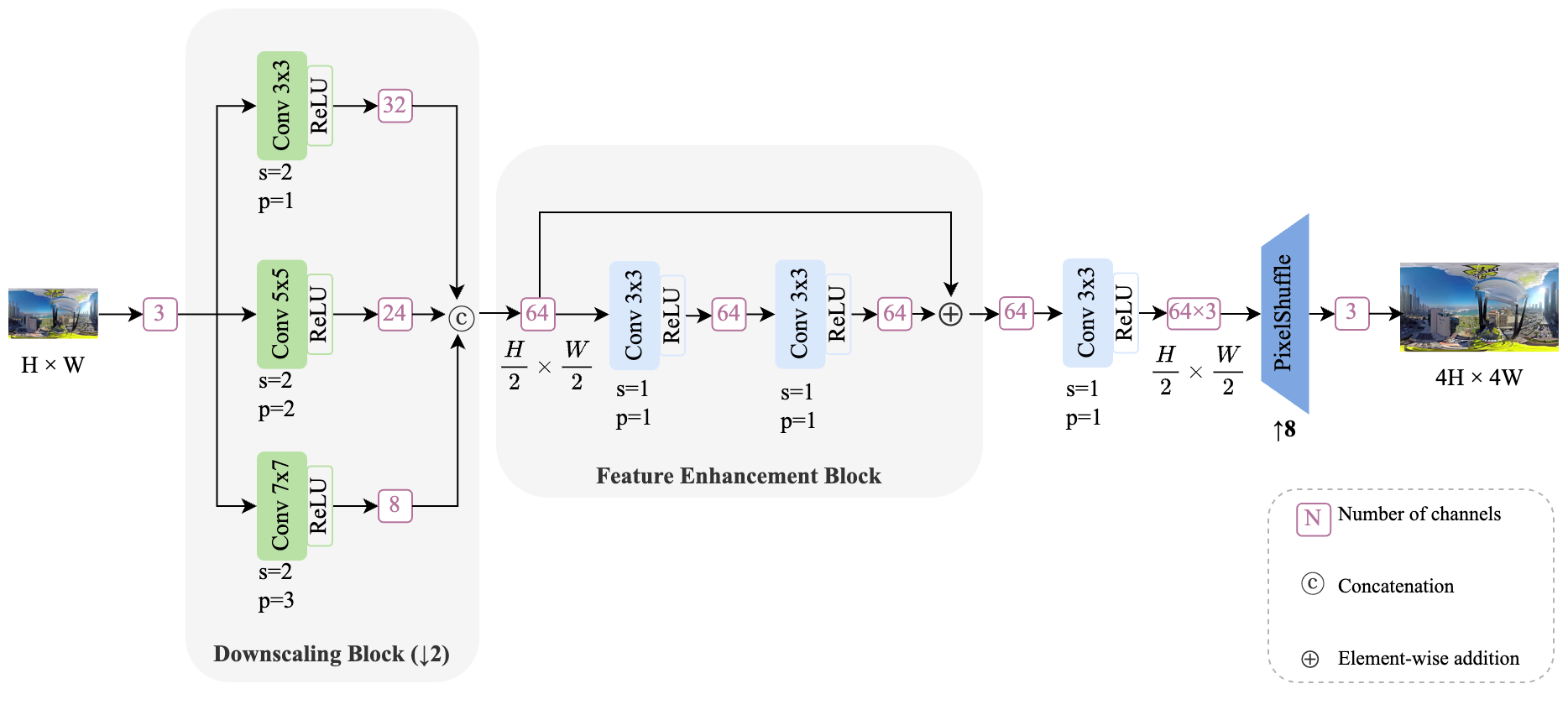}
    \caption{Team ATHENA: The overall architecture of the proposed SR network at the upscale factor $\times$4. The s and p parameters denote the stride and padding of the convolutional layers, respectively.}
    \label{fig:athena_network}
\end{figure*}

\vspace{1em}
\noindent \textbf{Network Architecture.} 
The ATHENA team proposed a lightweight and efficient \ac{cnn} for 360-degree video \ac{sr}, as depicted in Fig.~\ref{fig:athena_network}. The architecture and design choices are inspired by efficient \ac{sisr} methods~\cite{bilecen2023bpp, conde2023rtsr, ignatov2022efficient} and are based on extensive experimentation. The key points of the proposed network are as follows: ({\it i}) A downscaling block is introduced to reduce the image resolution while increasing the channel dimension. This block consists of three parallel pathways, each employing convolutional kernels of varying sizes to extract multi-scale features. The input images are downscaled by half with these strided convolutions, significantly decreasing the number of operations. ({\it ii}) The concatenated feature map is subsequently processed by two convolutional layers with residual connections, facilitating further feature propagation. ({\it iii}) Finally, another convolutional layer is applied before upscaling to the desired output resolution using pixel-shuffle. \\\\
\textbf{Implementation Details.}
The model was trained from scratch. \Ac{lr} patches of dimensions 64$\times$64 were randomly cropped from \ac{hr} images, with a mini-batch size of \num{4}. Data augmentation was performed on the input patches through random flipping and rotations. The network was trained by minimizing the Charbonnier loss using the Adam optimizer. The learning rate was initialized at \num{1e-4} and decayed by a factor of \num{0.1} every \num{1d4} epochs. The total number of training iterations was \num{5d5}.

\section{Experimental results}
\label{sec:results}

\subsection{Experimental setup}
\vspace{1em}
\subsubsection{Baselines} To benchmark the proposed methods on the ODVista dataset, we compare their performance against a diverse set of established \ac{sr} techniques. This comparative analysis includes both conventional interpolation methods and state-of-the-art \ac{ml}-based approaches, as detailed below: \\\\
\textbf{Bicubic interpolation~\cite{keys1981cubic}.} Bicubic interpolation, a widely employed method for image scaling in conventional \ac{sr} techniques, calculates the values of new pixels by applying a weighted average to the 16 neighboring pixels within a 4$\times$4 neighborhood. \\\\
\textbf{Lanczos filter~\cite{duchon1979lanczos}.} The Lanczos filter represents a more advanced conventional \ac{sr} method. It employs a sinc-based kernel, known as the Lanczos kernel, to estimate pixel values. Unlike bicubic interpolation, which operates within a 4$\times$4 pixel grid, the Lanczos filter can incorporate a larger number of surrounding pixels, with the specific count determined by the kernel size (e.g., Lanczos-3, Lanczos-4). In our implementation, we utilized a kernel size of 4. \\\\
\textbf{FSRCNN \cite{10.1007/978-3-319-46475-6_25}.} \Ac{fsrcnn} is a \ac{cnn}-based method, an evolution of \ac{srcnn}~\cite{dong2015image}, designed for real-time \ac{sr} applications. It features a streamlined architecture that processes \ac{lr} inputs directlybased, thereby reducing computational complexity. The notable performance of \ac{fsrcnn} can be attributed to its strategic use of deconvolution layers towards the end of the network, enabling efficient image upscaling and reducing processing time. \\\\
\textbf{SwinIR \cite{liang2021swinir}.} 
SwinIR, standing for Image Restoration using Swin Transformer, is a transformer-based model designed for diverse tasks, including \ac{sr}, image denoising, and compression artifact reduction~\cite{liu2021swin}. Building upon the Swin Transformer architecture~\cite{liu2021swin}, SwinIR processes images at multiple scales through a distinctive shifted windowing scheme for self-attention. This design empowers SwinIR to detect and interpret complex image patterns and long-range dependencies more effectively than traditional convolutional methods. \\\\
\textbf{RT4KSR \cite{zamfir2023towards}.} RT4KSR, an efficient state-of-the-art \ac{sr} method, employs a shallow convolutional architecture, a reparameterizable residual block, a high-frequency extraction block, and a pixel shuffle module. It achieves performance comparable to previous leading methods while maintaining real-time processing capabilities for \ac{hr} images. 
Notwithstanding

We intended to evaluate several state-of-the-art \ac{vsr} models, including BasicVSR++~\cite{chan2022basicvsr++}. However, the substantial memory footprint required for processing 4K resolution videos with 100 frames presented a significant challenge. Furthermore, the inherent computational complexity of \ac{vsr} methods led to prohibitive processing times. Consequently, we opted to utilize the aforementioned \ac{sisr} methods, recognizing their greater suitability for the real-time constraints of our evaluation framework. 
\subsubsection{Evaluation metrics} 

{\textbf{Perceptual quality metrics:} Perceptual quality metrics are essential for evaluating \ac{qoe} and enhancing the overall user experience in visual media systems. Significant efforts have been made in both academia and industry to design and refine these metrics. Cheon~\etal~\cite{cheon2021perceptual} provided a comprehensive survey of perceptual image quality assessment (IQA), reviewing both subjective and objective approaches, widely used benchmark datasets, evaluation protocols, and methodological advances. Similarly, Zhai~\etal~\cite{zhai2020perceptual} offered an extensive overview of video quality assessment (VQA), covering general-purpose and application-specific metrics across domains such as streaming, user-generated content and high frame rate video.}

{Among general-purpose methods, Min et al. in ~\cite{min2017blind, min2018blind} introduced the concept of Pseudo-Reference Image (PRI) and proposed a PRI-based blind IQA framework (BPRI), which generates a reference-like image from the distorted image by intentionally applying strong degradations. For 360-degree content, MC360IQA~\cite{sun2019mc360iqa} utilized a multi-channel CNN and viewport-based representations to capture distortions in omnidirectional imagery. More recently, Zhou et al.~\cite{zhou2023perception} proposed a U-shaped transformer-based IQA framework tailored for no-reference 360-degree image quality assessment.}

{However, in this work, our evaluation focused on WS-PSNR~\cite{sun2016ws} and WS-SSIM~\cite{zhou2018weighted} to ensure consistency with prior video super-resolution benchmarks and maintain objective comparability across submissions. To further assess quality in the context of bandwidth-constrained scenarios, we also computed the \ac{bd-br}~\cite{bjoentgaard2001} to quantify bitrate savings relative to the bicubic interpolation baseline across the four considered bitrate levels.}

\textbf{Runtime and computational efficiency evaluation:} Recognizing the critical importance of computational complexity in developing efficient \ac{sr} methods, we measured the runtime performance on a PC equipped with an Intel® Xeon 8280 CPU @ 2.70GHz (56 cores), 128GB RAM, and a 48GB VRAM NVIDIA RTX 6000 Ada graphics card. Furthermore, to offer a comprehensive understanding of the computational demands and efficiency of each method, we provided insights into the number of \acp{flop} and parameters involved.

\textbf{Scoring metric:} In addition, to comprehensively evaluate the trade-off between quality enhancement and runtime, we introduced a novel scoring metric, $Q$, that considers both factors. This metric, defined in \eqref{eq:q}, enables a balanced assessment of each method's performance, allowing us to rank them based on their overall effectiveness in improving video quality while maintaining computational efficiency.
\begin{figure}[pos=t]
\centering
 \includegraphics[width=0.99\linewidth]{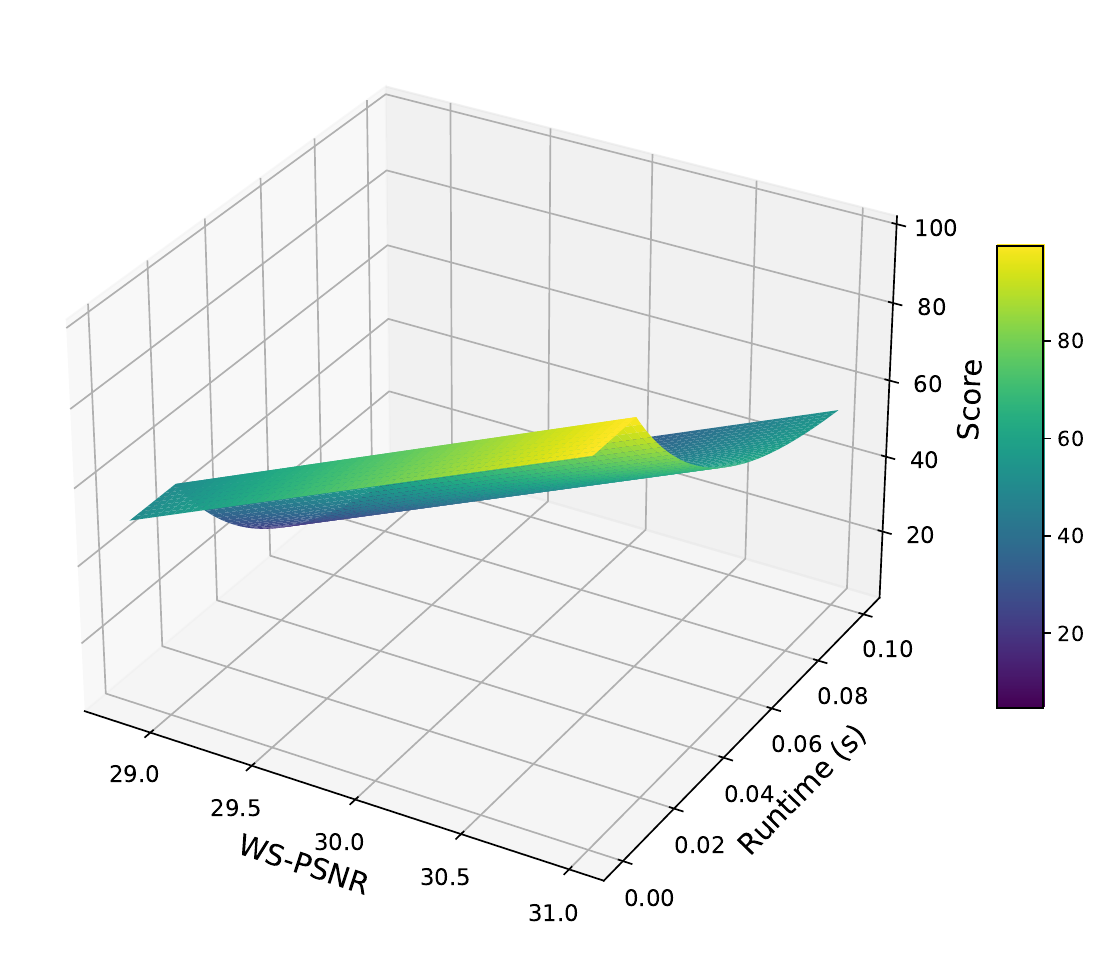}
\caption{3D Visualization of score metric $Q$ variation with \acs{ws-psnr} (dB) and runtime (s), with $\beta$ = \qty{0.5}, $\acs{ws-psnr}_{min}$ = \qty{28.8}{\decibel} and $\acs{ws-psnr}_{max}$ = \qty{31}{\decibel}.}
\label{score}
\end{figure}

\begin{equation}
\label{eq:q}
Q = (\beta \times \hat{Q} + (1 - \beta) \times C) \times 100,
\end{equation}
where $\beta$ serves as a weighting parameter (set to 0.5 in our evaluation), $\hat{Q}$ represents the normalized \ac{ws-psnr} score, and $C$ denotes the runtime evaluation score. The normalized quality score $\hat{Q}$ is computed as defined in \eqref{eq:normQ}.
\begin{equation}
\label{eq:normQ}
 \hat{Q} =   \frac{\ac{ws-psnr}-\ac{ws-psnr}_{min}}{\ac{ws-psnr}_{max} -\ac{ws-psnr}_{min}},
\end{equation}
{where $\text{WS-PSNR}_{\min}$ represents the minimum \ac{ws-psnr} value achieved by the least performing model (\ie Bicubic), and $\text{WS-PSNR}_{\max}$ denotes the theoretical maximum \ac{ws-psnr} values we consider in our evaluation (\qty{30}{\decibel} for a scaling factor of $\alpha=4$ and \qty{31}{\decibel} for $\alpha=2$). The runtime evaluation metric assigns a full score to models achieving a processing time of 0.016 seconds or less per 2K frame, as this speed enables a smooth 60 frames per second (fps) output, critical for high-quality live video streaming. To further emphasize the importance of real-time performance, we apply penalties to the final score for models with less than 60 fps. The slower the model, the larger the penalty, as expressed in~\eqref{eq:C}.}
\begin{equation}
\label{eq:C}
C =
  \begin{cases}
    1  & \quad \text{runtime } \leq 0.016, \\
   e^{B \times (0.016 -\text{runtime})}  & \quad \text{otherwise,} \text{ with } B=30.
  \end{cases}
\end{equation}
Fig.~\ref{score} illustrates the variation of the composite score, $Q$, as a function of runtime and \ac{ws-psnr}. Models achieving the highest scores are those that meet the real-time performance criterion of 60 fps while simultaneously delivering the highest \ac{ws-psnr} values.  The score is penalized when the runtime exceeds the real-time threshold or when the model's quality enhancement over the bicubic baseline method is marginal.

\subsection{Results and analysis}

\begin{table*}[]
\centering

\caption{Quantitative results of the 360-Degree Video Super Resolution and Quality Enhancement Challenge on the Test set.}

\label{tab:results}
\begin{tabular}{@{}
>{\columncolor[HTML]{FFFFFF}}l 
>{\columncolor[HTML]{FFFFFF}}l 
>{\columncolor[HTML]{FFFFFF}}c 
>{\columncolor[HTML]{FFFFFF}}c 
>{\columncolor[HTML]{FFFFFF}}c 
>{\columncolor[HTML]{FFFFFF}}c 
>{\columncolor[HTML]{FFFFFF}}c 
>{\columncolor[HTML]{FFFFFF}}c 
>{\columncolor[HTML]{FFFFFF}}c @{}}
\toprule
\cellcolor[HTML]{FFFFFF} & \cellcolor[HTML]{FFFFFF} & \cellcolor[HTML]{FFFFFF} & \cellcolor[HTML]{FFFFFF} & \cellcolor[HTML]{FFFFFF} & \cellcolor[HTML]{FFFFFF} & \cellcolor[HTML]{FFFFFF} & \cellcolor[HTML]{FFFFFF} & \cellcolor[HTML]{FFFFFF} \\
\multirow{-2}{*}{\cellcolor[HTML]{FFFFFF}Scale} & \multirow{-2}{*}{\cellcolor[HTML]{FFFFFF}Methods} & \multirow{-2}{*}{\cellcolor[HTML]{FFFFFF}WS-PSNR (dB) $\uparrow$} & \multirow{-2}{*}{\cellcolor[HTML]{FFFFFF}WS-SSIM $\uparrow$} & \multirow{-2}{*}{\begin{tabular}[c]{@{}c@{}}Runtime/\\ 2k frame (s) $\downarrow$\end{tabular}} & \multirow{-2}{*}{\cellcolor[HTML]{FFFFFF}Score Q $\uparrow$} & \multirow{-2}{*}{\begin{tabular}[c]{@{}c@{}}BD-BR VS\\ Bicubic $\downarrow$\end{tabular}} & \multirow{-2}{*}{\cellcolor[HTML]{FFFFFF}G-FLOPs $\downarrow$} & \multirow{-2}{*}{\cellcolor[HTML]{FFFFFF} \# Parameters $\downarrow$} \\ \midrule
\rowcolor[HTML]{90EE90} $2\times$ & \textbf{VACV}{\color{red}$ ^\maltese$} & 29.589 & 0.8217 & 0.0057 & \textbf{81.25} & -15.32\% & 45.641 & 315.120K \\
$2\times$ & ATHENA{\color{red}$ ^\maltese$} & 29.422 & 0.8177 & \textbf{0.0009} & 79.03 & -14.62\% & \textbf{15.340} & 105.216K \\
$2\times$ & IVCL{\color{red}$ ^\maltese$} & 29.645 & 0.8219 & 0.0298 & 58.26 & -16.91\% & 122.297 & 212.652K \\
$2\times$ & FFCIR{\color{red}$ ^\maltese$} & \textbf{29.761} & \textbf{0.8224} & 0.0420 & 45.61 & \textbf{-17.85\%} & 216.608 & 383.124K \\ 
\addlinespace[0.3ex] \hdashline \addlinespace[0.3ex]
$2\times$ & FSRCNN{\color{red}$ ^\maltese$} & 29.280 & 0.8149 & 0.0042 & 77.14 & -9.88\% & 38.102 & \textbf{24.683K} \\
$2\times$ & RT4KSR{\color{red}$ ^\maltese$} & 29.367 & 0.8192 & 0.0046 & 78.29 & -14.11\% & 22.423 & 151.992K \\
$2\times$ & SwinIR{\color{red}$ ^\maltese$} & 29.761 & 0.8250 & 1.5232 & 13.53 & -18.01\% & 544.667 & 910.152K \\
$2\times$ & Lanczos{\color{green} $^\clubsuit$} & 28.797 & 0.8110 & \xmark & \xmark & -1.04\% & \xmark & \xmark \\ 
$2\times$ & Bicubic{\color{green} $^\clubsuit$} & 28.743 & 0.8117 & \xmark & \xmark & Anchor & \xmark & \xmark \\ \midrule
\rowcolor[HTML]{90EE90} $4\times$ & \textbf{FFCIR}{\color{red}$ ^\maltese$} & \textbf{29.083} & \textbf{0.8090} & 0.0120 & \textbf{79.25} & \textbf{-42.82\%} & 55.857 & 394.824K \\
$4\times$ & IVCL{\color{red}$ ^\maltese$} & 28.920 & 0.8061 & 0.0050 & 75.57 & -38.78\% & 33.409 & 232.128K \\
$4\times$ & VACV{\color{red}$ ^\maltese$} & 28.918 & 0.8047 & 0.0058 & 75.52 & -38.76\% & 45.641 & 315.120K \\
$4\times$ & ATHENA{\color{red}$ ^\maltese$} & 28.425 & 0.7960 & \textbf{0.0004} & 64.37 & -26.77\% & \textbf{6.858} & 188.160K \\ 
\addlinespace[0.3ex] \hdashline \addlinespace[0.3ex]
$4\times$ & FSRCNN{\color{red}$ ^\maltese$} & 28.317 & 0.7912 & 0.0010 & 61.92 & -12.51\% & 33.334 & \textbf{24.683K} \\
$4\times$ & RT4KSR{\color{red}$ ^\maltese$} & 28.602 & 0.7991 & 0.0032 & 68.37 & -32.97\% & 6.739 & 183.240K \\
$4\times$ & SwinIR{\color{red}$ ^\maltese$} & 29.065 & 0.8099 & 0.4458 & 28.85 & -41.12\% & 136.167 & 929.628K \\
$4\times$ & Lanczos{\color{green} $^\clubsuit$} & 27.795 & 0.7814 & \xmark & \xmark & -3.96\% & \xmark & \xmark \\ 
$4\times$ & Bicubic{\color{green} $^\clubsuit$} & 27.790 & 0.7831 & \xmark & \xmark & Anchor & \xmark & \xmark \\
\bottomrule
\end{tabular}
{\begin{flushleft}
 {\color{red}$ ^\maltese$} \acs{ml}-based \acs{sr} methods, {\color{green} $^\clubsuit$} Handcrafted-based \acs{sr} methods.
   \end{flushleft}}
\end{table*}
Table~\ref{tab:results} presents the performance of the proposed methods and baseline models for both the 2$\times$ and 4$\times$ scaling tracks across several key metrics: quality enhancement (\ac{ws-psnr}, \ac{ws-ssim}), efficiency (runtime, \ac{flop}, and number of model parameters), and rate reduction (\acs{bd-br}). In addition, the $Q$ score, which quantifies the trade-off between quality enhancement and efficiency, is provided and serves as a basis for ranking the models. \\\\
{\bf Track 1 (2$\times$scaling).}
In Track 1, the proposed methods exhibited competitive performance across various metrics. Team VACV emerged as the top performer, achieving the highest $Q$ score of \qty{81.25}, demonstrating a favorable balance between quality enhancement (\ac{ws-psnr} of \qty{29.59}{\decibel}) and efficiency (runtime under \num{6} ms/2K frame), enabling a significant bandwidth savings of \qty{15.32}{\percent} compared to the bicubic baseline. Team ATHENA secured the second position with a $Q$ score of \qty{79.03}, a \ac{ws-psnr} of \qty{25.589}{\decibel}, and a \ac{bd-br} gain of \qty{14.62}{\percent}. Notably, the ATHENA team achieved the lowest runtime of less than \num{1} millisecond/2K frame, highlighting its exceptional efficiency. Teams IVCL and FFCIR achieved substantial quality enhancements with \ac{ws-psnr} values of \qty{29.645}{\decibel} and \qty{29.761}{\decibel}, respectively. However, their runtime performance fell short of the desired \num{60} \ac{fps} target, leading to their third and fourth rankings in this track.



\begin{figure*}[pos=t]
\centering
\begin{subfigure}[t]{0.48\linewidth}
    \centering
    \includegraphics[width=\linewidth]{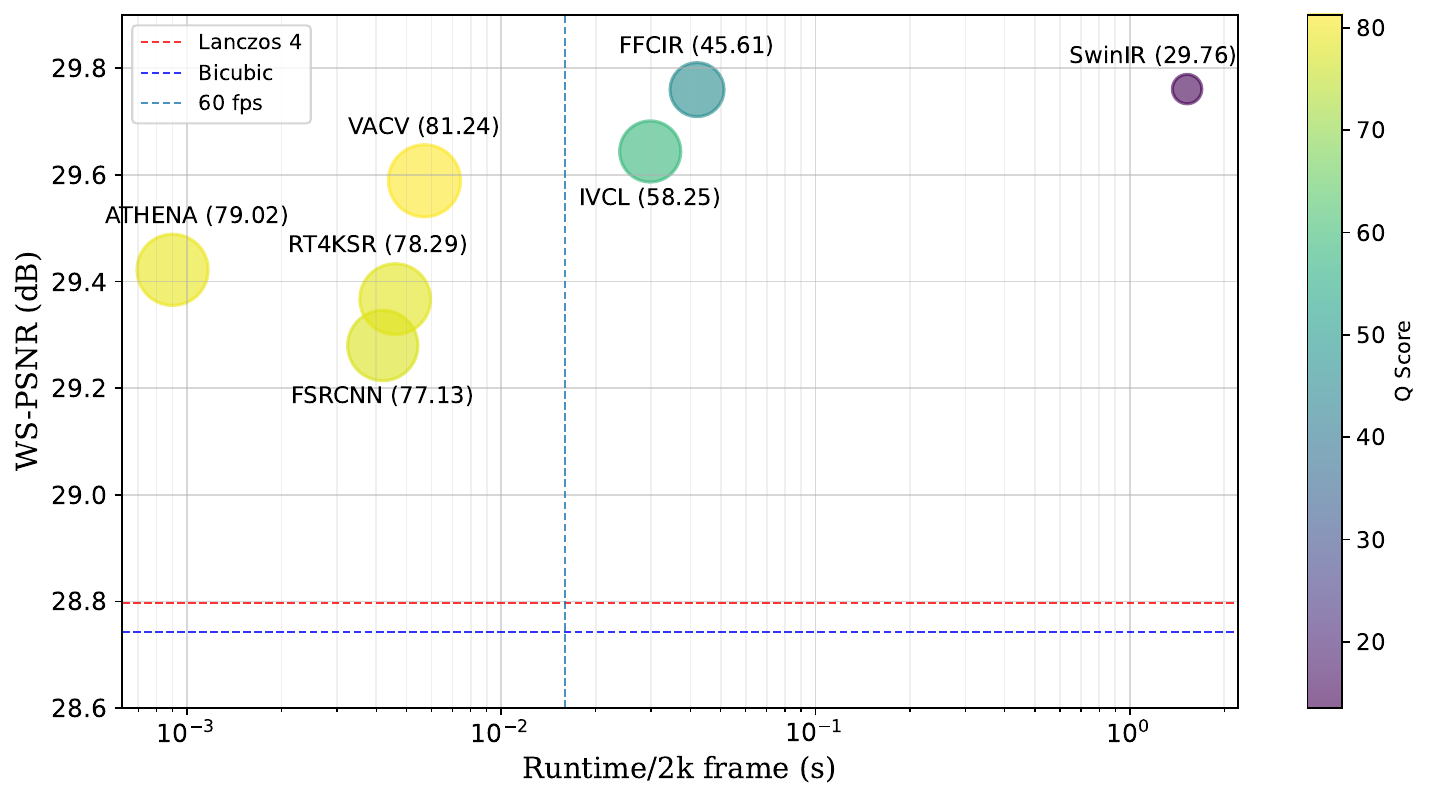}
    \caption{Track 1.}
    \label{psnr_vs_runtime_track1}
\end{subfigure}
\hfill
\begin{subfigure}[t]{0.48\linewidth}
    \centering
    \includegraphics[width=\linewidth]{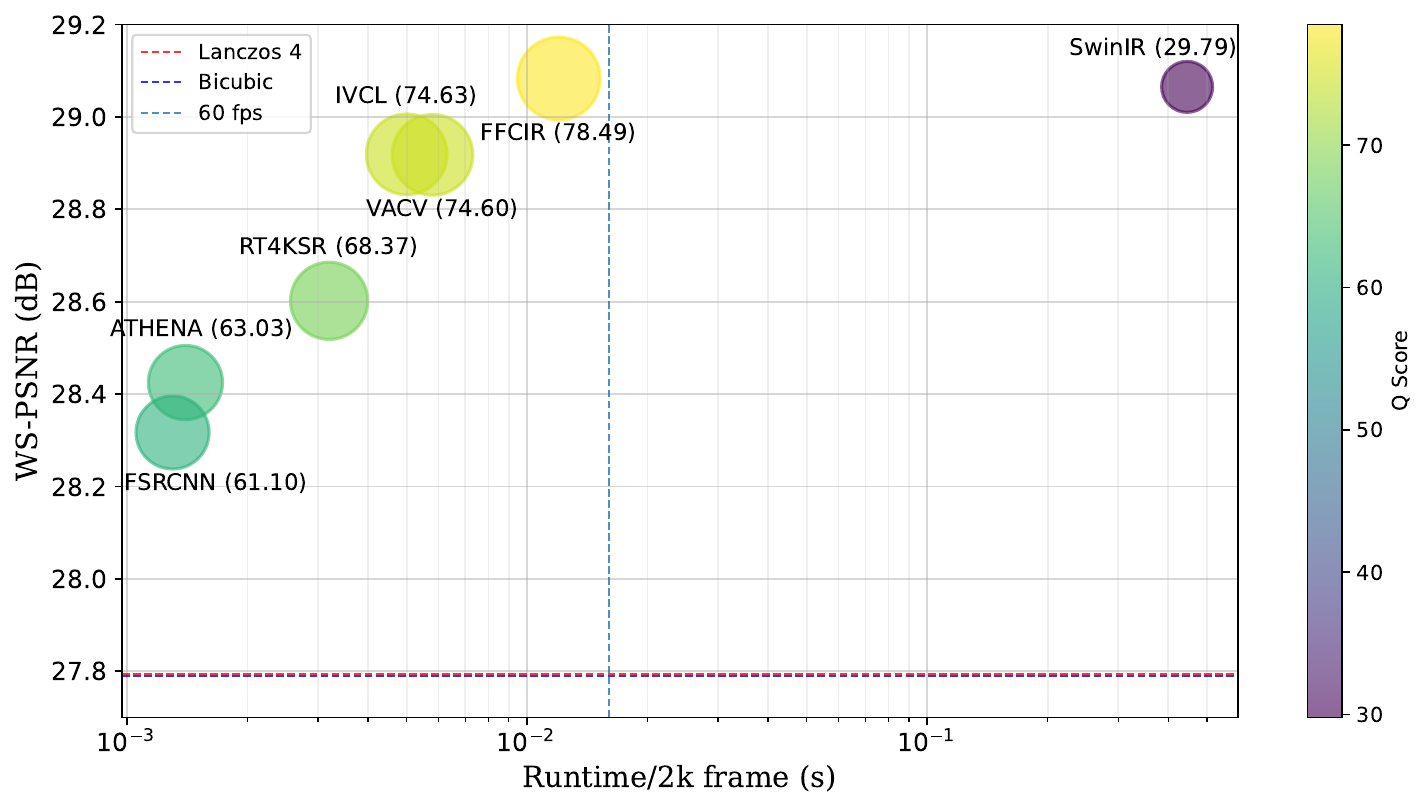}
    \caption{Track 2.}
    \label{psnr_vs_runtime_track2}
\end{subfigure}

\caption{Comparison of WS-PSNR versus Runtime for proposed methods and baseline models in two tracks.}
\label{fig:psnr_vs_runtime_combined}
\end{figure*}

\begin{figure*}[pos=t]
  \centering
  \includegraphics[width=1\linewidth]{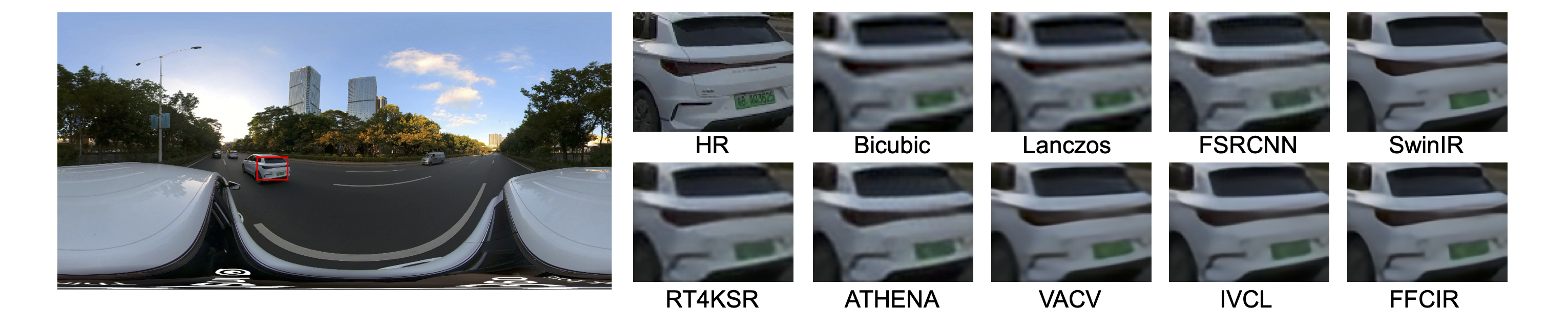}
\caption{Visual comparison of proposed methods and baseline models for $4\times$ scaling at a 2 Mbit/s compression bitrate.}
\label{SR_comparision}
\end{figure*}

Among the baseline methods, both \ac{fsrcnn} and RT4KSR significantly improve the quality compared to conventional approaches (Bicubic and Lanczos). However, they lagged behind all proposed methods in terms of overall performance, despite their runtime being less than \num{5} milliseconds/2k frame. SwinIR, while offering excellent quality, exhibited higher complexity, requiring an average of \qty{1.5} seconds to process a single 2K resolution frame. Fig.~\ref{psnr_vs_runtime_track1} illustrates the trade-off between quality (\ac{ws-psnr}) and runtime for the tested methods. \\\\
{\bf Track 2 (4$\times$scaling).} 
In Track 2, the FFCIR team emerges as the top performer, achieving the highest  score ($Q$) of \qty{79.25}, along with the best \ac{ws-psnr} score of \qty{29.083}{\decibel} and \ac{ws-ssim} of \qty{0.8090}. This results in an impressive bandwidth gain of \qty{42.82}{\percent} compared to the conventional method, all within a runtime of \num{12} milliseconds/2K frame. Following closely, the IVCL team secures the second position with a $Q$ score of \qty{75.57} and a \ac{ws-psnr} of \qty{28.920}{\decibel}, showcasing solid performance with an impressive inference time of \num{5} milliseconds. Team VACV trails IVCL slightly, achieving a $Q$ score of \qty{75.52} and a \ac{ws-psnr} of \qty{28.918}{\decibel}. Team ATHENA, while demonstrating a very impressive low inference time of \qty{0.5} milliseconds/2K frame, experiences a performance decrease with a $Q$ score of \qty{64.37}, caused by a lower quality enhancement performance.

Among the baseline methods, SwinIR exhibits a significant quality enhancement with a WS-PSNR of \qty{29.065}{\decibel}, but at the cost of high computational complexity, requiring \qty{0.45} seconds to process a single frame. In contrast, both \ac{fsrcnn} and RT4KSR improve the quality of restored frames compared to conventional methods (with gains of \qty{0.527}{\decibel} and \qty{0.812}{\decibel} respectively, compared to the Bicubic method), while maintaining real-time inference times of \num{1} millisecond and \qty{3.2} milliseconds, respectively.

Fig.~\ref{SR_comparision} provides a visual comparison between the tested methods. Teams FFCIR, IVCL, VACV, and SwinIR effectively restore high-frequency details and reduce blurring and compression artifacts, resulting in sharper and more natural edges. On the other hand, RT4KSR and the ATHENA team produce slightly blurry images, although they still outperform conventional methods significantly. Fig.~\ref{psnr_vs_runtime_track2} illustrates the trade-off between quality enhancement (\ac{ws-psnr}) and runtime for the tested methods. \\\\
{\bf Architecture Design Analysis.}
The proposed architectures enhance the performance of \ac{sr} models in mitigating low bitrate compression artifacts and improving overall quality while maintaining real-time processing capabilities. All proposed models adhere to a simple yet efficient single-frame restoration approach to prioritize both quality enhancement and runtime efficiency, as most \ac{vsr} methods tend to be computationally expensive. Each designed method incorporates a shallow feature extraction block with small convolutional filters, coupled with a depth-to-space (pixel shuffle) module at the output layer for upscaling the input frames. Two of the proposed methods, from teams FFCIR and VACV, are built upon the SAFMN architecture~\cite{SAFMN_2023_ICCV}, while team IVCL employs a Cross Stage Partial Network (CSPNet)~\cite{wang2020cspnet}. In contrast, team ATHENA utilizes shallow convolutional layers as the foundation of their approach.


\section{Conclusion}
\label{sec:conclusion}
In this paper, we presented the 360-Degree Video Super-Resolution and Quality Enhancement Challenge, aimed at addressing the critical challenges associated with low-bitrate streaming of \acp{odv}. We provided a comprehensive overview of the proposed challenge framework, encompassing two tracks focused on enhancing \ac{odv} resolution by factors of 2$\times$ and 4$\times$. The participating teams submitted solutions that employed single-frame restoration approaches, effectively balancing quality improvement with runtime efficiency. To mitigate the complexity inherent in incorporating temporal aspects, most models utilized shallow architectures characterized by small convolutional filters and channel sizes. Additionally, the depth-to-space (pixel shuffle) operation was favored over transposed convolutions to reduce computational overhead during image upsampling. Our rigorous evaluation, conducted on the ODVista dataset, benchmarked these methods against a variety of established \ac{sr} techniques, encompassing both traditional interpolation methods and state-of-the-art \ac{ml}-based approaches. The results underscore the efficacy of the proposed methods in enhancing video quality while adhering to real-time processing constraints. This collective effort significantly advances the field of super-resolution for \ac{odv}, establishing new benchmarks and paving the way for future research endeavors. {However, despite the encouraging results, the challenge and its evaluated methods are not without limitations. Most submitted solutions relied on single-frame architectures, which, while efficient, may underperform under complex motion or severe temporal distortions. Additionally, performance under extreme compression, such as ultra-low bitrates below 0.25 Mbps, remains a challenge and was not thoroughly explored in this edition. Future iterations may include more diverse artifacts, such as heavy compression, motion blur, or sensor noise, to better assess model robustness and generalization in real-world conditions.}

\balance
\bibliographystyle{model1-num-names.bst}
\bibliography{IEEEabrv}
\end{document}